\documentclass[fleqn,usenatbib]{mnras}
\usepackage{newtxtext,newtxmath,times,verbatim}

\usepackage[T1]{fontenc}
\DeclareRobustCommand{\VAN}[3]{#2}
\let\VANthebibliography\thebibliography
\def\thebibliography{\DeclareRobustCommand{\VAN}[3]{##3}\VANthebibliography}
\usepackage{graphicx}	% Including figure files
\usepackage{amsmath}	% Advanced maths commands
\usepackage{amssymb}	% Extra maths symbols
\usepackage{ulem}

\title[Formation of extremely massive disc galaxies]{Extremely massive disc galaxies in the nearby Universe form through gas-rich minor mergers}

\author[R. A. Jackson et al.]{R. A. Jackson,$^{1}$\thanks{E-mail: r.jackson@yonsei.ac.kr} S. Kaviraj,$^{2}$ G. Martin, $^{3,4}$ J. E. G. Devriendt,$^{5}$ E. A. Noakes-Kettel,$^{2}$ \newauthor 
J. Silk,$^{5,6,7}$ P. Ogle $^{8}$ and Y. Dubois$^{6}$ \\
$^{1}$Department of Astronomy and Yonsei University Observatory, Yonsei University, Seoul 03722, Republic of Korea\\
$^{2}$Centre for Astrophysics Research, School of Physics, Astronomy and Mathematics, University of Hertfordshire, Hatfield, AL10 9AB, UK\\
$^{3}$Korea Astronomy and Space Science Institute, 776 Daedeokdae-ro, Yuseong-gu, Daejeon 34055, Korea\\
$^{4}$Steward Observatory, University of Arizona, 933 N. Cherry Ave, Tucson, AZ 85719, USA\\
$^{5}$Dept of Physics, University of Oxford, Keble Road, Oxford OX1 3RH UK\\
$^{6}$Institut d'Astrophysique de Paris, Sorbonne Universit\'es, UMPC Univ Paris 06 et CNRS, UMP 7095, 98 bis bd Arago, 75014 Paris, France\\
$^{7}$Department of Physics \& Astronomy, The Johns Hopkins University, Baltimore, MD 21218, USA\\
$^{8}$Space Telescope Science Institute, 3700 San Martin Drive, Baltimore, MD 21218, USA\\
}

\begin{document}
\label{firstpage}
\pagerange{\pageref{firstpage}--\pageref{lastpage}}
\maketitle

\begin{abstract}
In our hierarchical structure-formation paradigm, the observed morphological evolution of massive galaxies -- from rotationally-supported discs to dispersion-dominated spheroids -- is largely explained via galaxy merging. However, since mergers are likely to destroy discs, and the most massive galaxies have the richest merger histories, it is surprising that any discs exist at all at the highest stellar masses. Recent theoretical work by our group has used a cosmological, hydrodynamical simulation to suggest that extremely massive (M$_*$ $>$ 10$^{11.4}$ M$_\odot$) discs form primarily via minor mergers between spheroids and gas-rich satellites, which create new rotational stellar components and leave discs as remnants. Here, we use UV-optical and HI data of massive galaxies, from the SDSS, GALEX, DECaLS and ALFALFA surveys, to test these theoretical predictions. Observed massive discs account for $\sim$13 per cent of massive galaxies, in good agreement with theory ($\sim$11 per cent). $\sim$64 per cent of the observed massive discs exhibit tidal features, which are likely to indicate recent minor mergers, in the deep DECaLS images (compared to $\sim$60 per cent in their simulated counterparts). The incidence of these features is at least four times higher than in low-mass discs, suggesting that, as predicted, minor mergers play a significant (and outsized) role in the formation of these systems. The empirical star-formation rates agree well with theoretical predictions and, for a small galaxy sample with HI detections, the HI masses and fractions are consistent with the range predicted by the simulation. The good agreement between theory and observations indicates that extremely massive discs are indeed remnants of recent minor mergers between spheroids and gas-rich satellites.

\end{abstract}

%.....................................................................


%.....................................................................

\section{Introduction}

Observational studies of the morphology of massive galaxies show that, while discs dominate the high-redshift Universe, the morphologies of nearby massive galaxies are mostly spheroidal in nature \citep{Bernardi2003,Wuyts2011,Ryan2012,Conselice2014,Buitrago2014,Shibuya2015}. In our standard hierarchical structure-formation paradigm, this morphological change, from rotationally-supported discs to dispersion-dominated spheroids, is largely explained by galaxy merging \citep{Toomre1977,Barnes1992,Bournaud2007,DiMatteo2007,Oser2010,Kaviraj2010,Kaviraj2011,Dubois2013,Dubois2016,Lofthouse2017,Welker2018,Martin2018b}. The large gravitational torques produced during merger events are capable of randomising the ordered rotational orbits of stars within the merging progenitors and creating dispersion-dominated systems \citep[e.g.][]{Springel2005,Hilz2013,Font2017,Martin2018b,Martin2019}. 

The role of merging is thought to become increasingly more important at higher stellar masses. In particular, significant merging activity is considered essential for galaxies to achieve the highest stellar masses \cite[e.g.][]{Faber2007,McIntosh2008,Cattaneo2011} beyond the knee of the galaxy mass function \citep[M$_*$ $\gtrsim$ 10$^{10.8}$ M$_\odot$;][]{Li2009,Kaviraj2017}, because star formation via direct gas accretion is no longer sufficient to drive the requisite stellar mass growth. However, since mergers can destroy discs, and the most massive galaxies have the richest merger histories, it is surprising that both observational \citep[e.g.][]{Conselice2006, Ogle2016, Ogle2019} and theoretical \citep[e.g.][]{Martin2018b,Jackson2020disc} studies suggest that a significant minority of galaxies at the highest stellar masses (M$_*$ $>$ 10$^{11.4}$ M$_\odot$) have discy morphologies.  

If discs exist in the stellar mass regime when mergers are frequent, their merger histories must involve peculiarities that either rejuvenate discy components or allow these discs to survive the mergers themselves. For example, theoretical work has shown that in gas-rich mergers, the gas brought in by the progenitors can create new discy stellar components in the remnants \citep[e.g.][]{Springel2005,Robertson2006,Governato2009,Hopkins2009,Font2017,Martin2018b,Peschken2019}. In \citet[][J20 hereafter]{Jackson2020disc} we have used Horizon-AGN, a cosmological hydrodynamical simulation \citep{Dubois2014,Kaviraj2017}, to probe the potential channels by which massive discs may form in the standard paradigm. J20 showed that extremely massive (M$_*$ $>$ 10$^{11.4}$ M$_\odot$) discs do exist in the simulation and are created via two channels. In the primary channel, which accounts for $\sim$70 per cent of these systems, a significant merger with a mass ratio greater than 1:10, between a massive spheroid and a gas-rich satellite, `spins up' the spheroid by creating a new rotational stellar component, and leaves a massive disc as the remnant. In the secondary channel, a system maintains a disc throughout its lifetime, due to an anomalously quiet merger history, with a merger rate that is more than a factor of two lower than in galaxies of a comparable stellar mass. This enables the galaxy to retain its gas reservoir more easily.

Several studies in the literature have also explored the formation mechanisms of massive disc galaxies. \citet{Martig2021} have looked at the formation history of NGC 5746 (an edge-on disc galaxy with M$_{\star}$ $\sim$ $10^{11}$M$_{\odot}$). They find that this galaxy formed its massive disc early and only underwent one significant merger ($\sim$1:10 mass ratio), without disruption to the disc component of the galaxy (consistent with one of the channels for massive disc formation outlined in J20). \citet{Ogle2016,Ogle2019} have suggested that the local massive discs in their study may have formed via major mergers (mass ratios greater than 1:4) between two gas-rich spiral galaxies. \citet{Monachesi2016} studied the outskirts of massive disc galaxies, finding variations in the median colours as a function of radius. They conclude that this is indicative of several small, accreted objects which have built up the outskirts of these galaxies (consistent with the minor-merger hypothesis in J20). \citet{Zeng2021} have used the Illustris-TNG simulation \citep{Nelson2019}, to show that disc galaxies in the mass range (M$_{\star} > 8\times10^{10}$ M$_{\odot}$) form via three channels. They find that $\sim8$ per cent of such discs have a quiescent merger history and remain discs throughout their lifetime, whilst $\sim54$ per cent have a significant increase in their bulge components before later becoming discs again. In their third channel, such discs experience prominent mergers but survive to remain disc-like. However, it is worth noting that this study includes many galaxies with significantly lower stellar mass than those in J20, which likely influences the creation channels and makes it difficult to directly compare the two studies. %Conversely, \citet{DiTeodoro2021} study a population of extremely massive spiral galaxies (with M$_{\star} > 10^{11}$). They use kinematics to study stellar/baryonic Tully-Fisher and Fall scaling relations, finding that these galaxies lie on these relations determined by lower mass galaxies. They conclude that this indicates these spirals are scaled-up versions of less massive discs.

If massive discs are indeed formed via the channels suggested in J20, the simulation makes some specific predictions that are testable. For example, J20 predicts that massive discs should comprise $\sim$11 per cent of the massive galaxy population. Since they are typically created via recent minor mergers, a large fraction of these galaxies should exhibit tidal features in deep optical images. Given the gas-rich nature of these mergers, J20 suggests that these systems should show reasonably high star-formation rates (SFRs) of a few solar masses per year. Finally, since massive ellipticals are predicted to have their last significant mergers at higher redshifts, and since these mergers are not gas-rich, both the fraction of tidally-disturbed galaxies and their SFRs should be elevated in the massive discs, compared to that in their spheroidal counterparts.

The purpose of this observational paper is to confront the theoretical predictions from J20 with multi-wavelength survey data, in order to establish whether the predictions are indeed supported by the observations. This paper is structured as follows. In Section \ref{sec:sample}, we describe the construction of a sample of massive galaxies and their multi-wavelength data, using the Sloan Digital Sky Survey \citep[SDSS;][]{York2000,Abazajian2009}, the Dark Energy Camera Legacy Survey \citep[DECaLS;][]{Dey2019}, the Galaxy Evolution Explorer \citep[GALEX;][]{Morrissey2007} and the Arecibo Legacy Fast ALFA \citep[ALFALFA;][]{Haynes2011} survey. We also describe the process of classifying galaxy morphology and identifying objects which have tidal features via visual inspection. In Section \ref{sec:properties}, we compare the observed properties of the massive discs in our sample with the predictions of J20. We summarise our findings in Section \ref{sec:summary}.\\

%.....................................................................

\section{A sample of massive galaxies in the nearby Universe}
\label{sec:sample}

We construct our sample of nearby massive galaxies from the SDSS using the MPA-JHU value-added catalog \footnote{https://wwwmpa.mpa-garching.mpg.de/SDSS/DR7/}. We select objects which have SDSS spectroscopic redshifts in the range $0.03<z<0.1$ and where the lower limit of their stellar masses, taken from the MPA-JHU catalog \citep{Brinchmann2004}, is greater than $10^{11.4}$ M$_{\odot}$. We identify 708 galaxies which match these criteria. 

The lower limit of the mass range ensures that our empirical sample of massive galaxies is comparable to those in the theoretical study of J20. The upper limit of the redshift range produces a sample with a median redshift that matches that of the simulated galaxies in J20. Standard depth SDSS imaging is typically too shallow to reveal faint tidal features from minor mergers \citep[e.g.][]{Kaviraj2010,Kaviraj2014b}. Instead, we use images from DECaLS, which provides deep optical imaging over $\sim$14,000 square degrees in the northern hemisphere. The DECaLS images in the $g$, $r$ and $z$ bands have 5$\sigma$ point-source depths of $\sim$24.0, $\sim$23.4 and $\sim$22.5 magnitudes respectively, roughly 1.5 magnitudes deeper than their SDSS counterparts, with similar seeing. At our redshifts of interest, the impact of surface-brightness dimming is still relatively minor, making it possible to see faint structures like tidal features in deep optical images \citep[e.g.][]{Kaviraj2019}.

\begin{figure*}
\centering
\includegraphics[width=\textwidth]{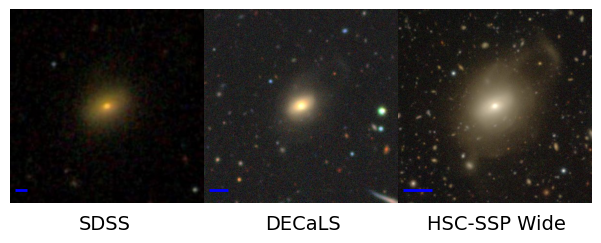}
\caption{Images of the same galaxy from the SDSS, DECaLS and the HSC-SSP Wide surveys. DECaLS and HSC-SSP are $\sim$1.5 and $\sim$4 mags deeper than the SDSS respectively. While the tidal features are invisible in the SDSS, they become progressively more easily visible in the deeper images. This figure highlights the need to use a deep-wide survey like DECaLS to detect galaxies that have recently undergone a merger. The blue bar in the bottom left indicates 10 arcseconds in each image. This figure may look better on screen than in print.}
\label{fig:depth_gals}
\end{figure*}

To demonstrate the improvement in our ability to detect merger-induced tidal features in deeper imaging, we show, in Figure \ref{fig:depth_gals}, images of the same galaxy from the SDSS, DECaLS and the Hyper Suprime-Cam Subaru Strategic Program \citep[HSC-SSP][]{Aihara2019}, which is a further $\sim$2 mags deeper than DECaLS. While the tidal features are invisible in the SDSS image, they are progressively more visible in the DECaLS and HSC-SSP images (and clearest in the deepest HSC-SSP image). Note that, while the HSC-SSP is the deepest wide-area optical survey currently available, its footprint ($\sim$1500 deg$^2$) is considerably smaller than that of DECaLS and therefore not suitable for this study. Indeed, only 4 out of the 708 massive galaxies in our sample are in the HSC-SSP footprint. 

\begin{figure*}
\centering
\includegraphics[width=\columnwidth]{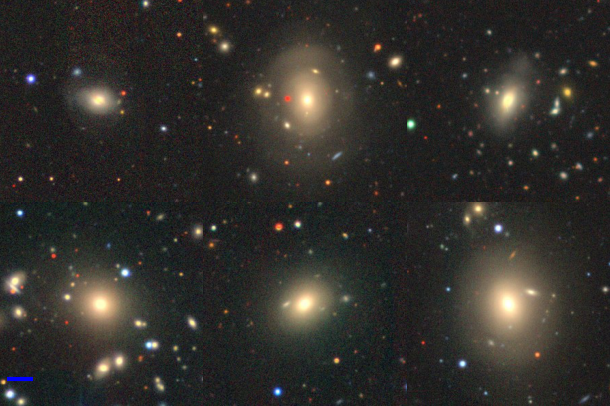}
\hspace{0.5cm}
\includegraphics[width=\columnwidth]{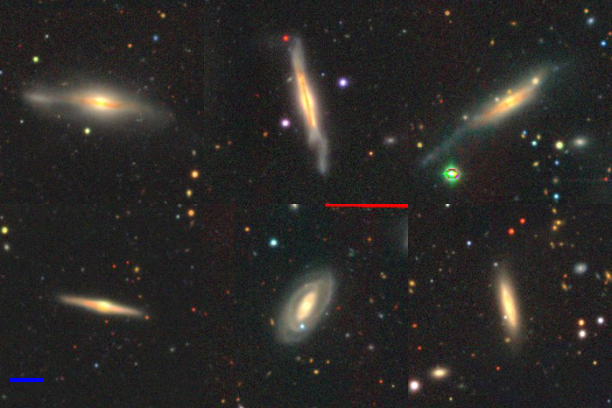}
\caption{Examples of spheroids (left) and discs (right) from our massive galaxy sample. The images are $grz$ composites from DECaLS. `Disturbed' galaxies (i.e. those that show evidence of tidal features such as tidal tails, plumes and prominent shells) are shown in the top row, while `relaxed' galaxies (i.e. those that do not show evidence of tidal features) are shown in the bottom row. The blue bar in the bottom left panel indicates 10 arcseconds in each image. This figure may look better on screen than in print.}
\label{fig:spirals}
\end{figure*}

\begin{figure*}
\centering
\includegraphics[width=0.965\textwidth]{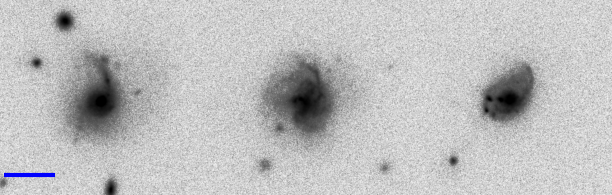}
\caption{Example $z$-band images of simulated massive disc galaxies (from J20) in the Horizon-AGN simulation. These images are made to match the DECaLS observations, including the surface brightness limit, pixel scale and point spread function. The simulated galaxies show tidal features, as is the case in their observed counterparts. The blue bar in the left panel indicates 25 kpc in each image.}
\label{fig:mock_gals}
\end{figure*}

We use SFRs from the GALEX-SDSS-WISE Legacy Catalog \citep[GSWLC;][]{Salim2016}, which are derived via spectral energy distribution (SED) fitting using total magnitudes from the GALEX and SDSS surveys. SFRs derived using total magnitudes are more likely to be representative of the total SFR of the system than those derived, for example, using emission lines measured within the SDSS fibre (which will only sample the star formation activity in the central regions of the galaxy). These total SFRs are more appropriate for comparison to the SFRs measured in simulated galaxies, where the entire galaxy is used to measure the SFR. Finally, we use HI masses from the ALFALFA survey. 

%.....................................................................

\subsection{Morphological classification and identification of merger-induced tidal features}

We visually inspect the composite $grz$ images of each individual massive galaxy from DECaLS to morphologically classify it as either a spheroid (which includes ellipticals and lenticular systems) or a disc. For each system, we also flag the presence of tidal features. In Figure \ref{fig:spirals}, we show a representative sample of galaxies, classified as spheroids (left-hand panel) and discs (right-hand panel) respectively. The top row of each panel shows examples of galaxies classified as `disturbed' (i.e. those which exhibit the presence of tidal features), while the bottom row shows galaxies classified as `relaxed' (i.e. those in which no tidal features are present).

%.....................................................................

\section{Observed properties of massive galaxies in the local Universe}
\label{sec:properties}

\subsection{Morphological properties and host halo mass}

We begin our analysis by considering the morphological properties of galaxies in our observational sample (summarised in Table \ref{tab:tidalfeat}). $\sim$13 per cent of our massive galaxies are discs, of which $\sim$64 per cent show evidence for tidal features (the corresponding value for early-types is $\sim$31 per cent). This is a lower limit because, as indicated by Figure \ref{fig:depth_gals}, it is likely that the fraction of galaxies with tidal features could be higher in deeper imaging. The corresponding fraction of systems in the low-mass disc population, which show tidal features in SDSS Stripe 82 images (which have similar depth and seeing to DECaLS), is between $\sim$11 and 17 per cent, depending on the specific morphology of the low-mass discs in question \citep[e.g. Sa/Sb/Sc/Sd, see][]{Kaviraj2014b}. The median stellar mass of the low-mass disc population in \citet{Kaviraj2014b} is $\sim$10$^{10.3}$ M$_{\odot}$ \citep[][]{Kaviraj2014a}. The fraction of tidal features in massive discs is therefore significantly elevated (by at least around a factor of 4) compared to their low-mass counterparts, implying that the formation of these galaxies involves a much higher incidence of recent mergers. 

\begin{center}
\begin{table}
\centering
\begin{tabular}{|| c | c | c |}
\hline
\hline

1 & 2 & 3\\
\hline
Morphology & Number fraction & Fraction with tidal features \\
 
\hline
\hline
Disc & 0.13 & 0.64\\
\\
Spheroid & 0.87 & 0.31\\
\\
\end{tabular}
\caption{The morphological properties of nearby massive (M$_*$ $>$ 10$^{11.4}$ M$_\odot$) galaxies. Columns: (1) morphological class (2) number fraction of galaxies in this morphological class (3) fraction of galaxies in this morphological class which exhibits tidal features in the DECaLS images. Massive discs show a high incidence of tidal features (which are significantly elevated, by at least around a factor of 4, compared to their low-mass counterparts), suggesting that the formation of these systems involves a high prevalence of mergers, consistent with the theoretical predictions in J20.}
\label{tab:tidalfeat}
\end{table}
\end{center}

To confirm that the tidal-feature fraction in the observed galaxies is indeed indicative of the formation methods predicted by J20, we also construct realistic mock images of the massive discs from J20 and estimate the corresponding tidal-feature fraction from the simulation (Figure \ref{fig:mock_gals}). We produce produce two sets of mock $z$-band images using the DECaLS pixel scale (0.262 arcseconds) and point spread function (1.10 arcseconds). The first imposes the DECaLS surface brightness limit \citep[$\sim$27.9 mag arcsec$^{-2}$,][]{Hood2018} while the second has no limit. 
%In Table \ref{tab:mockimages} we summarise the finding of this analysis. 
Visual inspection of these images shows that the surface-brightness-limited images exhibit a tidal feature fraction of $\sim$60 per cent, which is comparable to the fraction ($\sim$64 per cent) in the observed images. However, if the images without a surface-brightness limit are used the tidal feature fraction increases to $\sim$74 per cent, demonstrating that faint tidal features can be missed even at the depth of DECaLS (as illustrated in Figure \ref{fig:depth_gals})\footnote{A recent study by \citet{Blumenthal2020}, using mock SDSS images, has suggested that tidal features may not be readily visible from either major or minor mergers. It is important to note, however, that Blumenthal et al. consider galaxy pairs with a total stellar mass of 10$^9$ M$_{\odot}$ or greater. In our study, the galaxy pair has a mass of at least 10$^{11.4}$ M$_{\odot}$. Indeed, the satellites that are accreted by the larger progenitors in the events that create the massive discs are themselves likely to have stellar masses around 10$^{10.8}$ M$_{\odot}$ (see the analysis in Section \ref{sec:star formation rates}). Tidal features are brighter when the objects involved are more massive, simply because there is more material in the features \citep[e.g.][]{Kaviraj2014b}. Hence the Blumenthal et al. study, which will be dominated by mergers between much lower mass galaxies, as a result of the shape of the galaxy stellar mass function \citep[e.g.][]{Wright2017}, is not comparable to our work. As we note in our analysis above, tidal features are very common in galaxies in the mass range of interest in this study, in the Horizon-AGN simulation.}.
%Therefore not only are tidal features being missed, even at the depth of DECaLS, but the tidal feature fraction is similar to the rejuvenated disc fraction predicted in J20 when SB is no longer a limiting factor.

\begin{comment}
\begin{center}
\begin{table}
\centering
\begin{tabular}{|| c | c ||}
\hline
\hline

1 & 2\\
\hline
Tidal feature frac. (SB limited) & Tidal feature frac. (no limit) \\
 
\hline
\hline
0.60 & 0.74\\
\\
\end{tabular}
\caption{}
\label{tab:mockimages}
\end{table}
\end{center}
\end{comment}

The high frequency of tidal features in massive discs suggests that most of these interactions are unlikely to be major mergers (mass ratios greater than 1:4). This is because, in our mass range of interest, the major merger fraction at low redshift is only a few per cent \citep[e.g.][]{Darg2010,Mundy2017}, with the tidal features from such interactions remaining visible for 2-3 Gyrs at the depth of the DECaLS images \citep[e.g.][]{Mancillas2019}. The majority of these events are therefore likely to be minor mergers (mass ratios between 1:4 and 1:10)\footnote{It is worth noting that the tidal features seen around the massive discs are unlikely to be the result of fly-bys. All massive discs in the J20 study are the result of actual galaxy mergers and not fly-bys, which cannot produce the strong morphological transformation required for the creation of these systems. In terms of the presence of tidal features, while low-mass (M$_{\star}$$<$10$^{8}$ M$_{\odot}$) dwarf galaxies can sometimes produce tidal features as a result of very close fly-bys \citep[e.g.][]{Jackson2021}, these features already become rare in fly-by events involving relatively massive satellites with stellar masses around 10$^{9}$ M$_{\odot}$ \citep{Martin2021}. More massive satellites (such as the ones involved in the creation of the massive discs, see text in Section \ref{sec:star formation rates}) are therefore extremely unlikely to produce tidal features.}, which are typically 4-5 times more frequent than their major counterparts \citep[e.g.][]{Jogee2009,LopezSanjuan2010,Lotz2011,Kaviraj2014b,RodriguezGomez2015}. 

These observational results appear consistent with the theoretical predictions of J20, which suggest that $\sim$11 percent of the massive galaxies in the simulation are discs, and that these systems form via recent minor mergers that take place within the last $\sim$2-3 Gyrs. In addition, massive spheroids in the simulation typically underwent their most recent mergers at higher redshift than their discy counterparts. Given that the tidal features from these mergers will have had more time to fade, this will result in fainter tidal features at $z\sim0$, which appears consistent with the lower tidal fraction observed in the massive early-types in our observed sample. 

Since relatively dense neighbourhoods like groups are likely to favour mergers \citep[e.g.][]{Kaviraj2015}, we consider the local environments of our massive discs, both in the observed sample and their theoretical counterparts. J20 suggest that extremely massive discs usually inhabit relatively massive dark-matter halos (which are indicative of relatively dense environments, such as large groups or clusters). Therefore, if the observed sample is representative of the simulated population, they should also reside in such massive halos. To study the local environment of our observed massive discs, 
we use the galaxy group catalogue from \citet{Yang2007} which provides halo masses for our observed galaxies.

\begin{figure}
\centering
\includegraphics[width=\columnwidth]{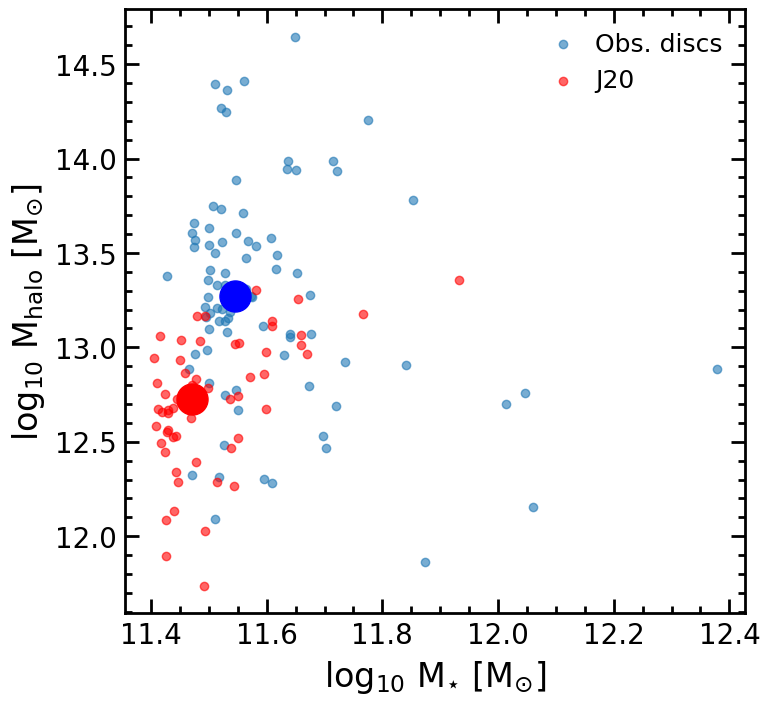}
\caption{Stellar mass vs dark matter halo mass for our observed massive discs and their theoretical counterparts in J20. The larger points show the median stellar and halo masses for both samples. The halo masses for the observed sample are taken from \citet{Yang2007}. The lack of simulated systems at very high stellar mass is due to the fact that the simulation volume is only $\sim$100 Mpc and therefore very rare objects, like the most massive galaxies, are absent from the simulation box. Despite the bias towards larger stellar masses in the observed sample, both populations of massive discs primarily reside inside massive halos, indicating relatively dense environments, which is consistent with the high incidence of tidal features in this population.}
\label{fig:halomass}
\end{figure}

Figure \ref{fig:halomass} shows stellar vs halo mass for our observed discs and compares them to their theoretical counterparts from J20. The paucity of simulated systems at very high stellar mass is driven by the fact that the simulation volume is only $\sim$100 Mpc and therefore very rare objects, like galaxies at the highest stellar masses, are absent from the simulation box. Both the observed and simulated galaxies inhabit large halos, with the median stellar and halo mass of both samples, denoted by the larger points, lying on a locus of increasing stellar and halo mass. To further this argument we compare how the halo masses of the massive discs compare to the distribution of halo masses of all galaxies with M$_{\star}>$10$^{10}$ M$_{\odot}$. We find that both the observed and simulated sample of massive discs occupy above the upper 10 percentile of halo masses. They therefore reside in very massive halos, consistent with the high frequency of tidal features seen in these systems.

%\textbf{We also consider the halo properties of the observed disc population compared to that of J20. In J20 it was shown that extremely massive discs usually dominate their respective group/cluster and therefore should inhabit massive halos. Therefore if the observed sample is representative of the simulated population, they should also reside in large halos. To study this we use the galaxy group catalogue from \citet{Yang2007} which includes halo masses for our observed galaxies. In Figure \ref{fig:halomass} we show the stellar vs halo mass for all the observed discs and those found in J20. Although the observed galaxies tend to higher stellar mass, most likely due to the error cut in the galaxy selection, they inhabit large halos just like the galaxies in J20. Indeed the median stellar and halo mass of both samples, denoted by the larger points, appear to lie on a locus of increasing stellar and halo mass. To further this argument we compare how the halo masses of these massive discs compare to the distribution of halo masses for galaxies with M$_{\star}>$10$^{10}$M$_{\odot}$. We find that both the observed and simulated sample of massive discs occupy above the 90th percentile of the distribution of halo masses. They therefore reside in very massive halos, which are likely to dominate their local environment.}

%.....................................................................

\subsection{Star formation rates and atomic gas properties}
\label{sec:star formation rates}

We proceed by considering the SFRs of our observed massive galaxies. %As shown in J20, massive discs in the Horizon-AGN simulation form via one of two mechanisms. The primary channel is a minor merger between an massive early-type galaxy and a gas-rich companion which rebuilds a disc, leaving a massive disc as the remnant. The secondary channel involves a galaxy that maintains its discy morphology (and relatively high gas fraction) throughout its history due to an anomalously quiet merger history. 
If the predicted formation mechanisms of massive discs in J20 are broadly accurate, then the expectation is that the SFRs in the observed and simulated massive discs should be comparable and that the SFRs in the discs should be significantly enhanced compared to that in the spheroids.

\begin{figure}
\centering
\includegraphics[width=\columnwidth]{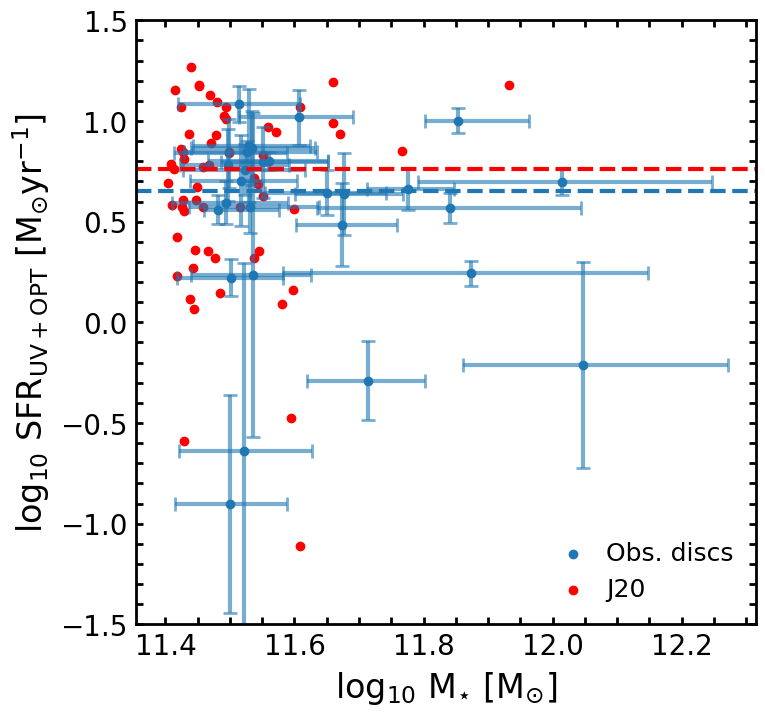}
\caption{SFR as a function of stellar mass for massive discs in our observational sample (blue) and the simulated sample of J20 (red). Dashed lines indicate median values. The SFRs from the observations and simulations show comparable median values and distributions. While we omit spheroids for clarity, the medians and distributions for all populations are summarised in Table \ref{tab:sfr}. The paucity of simulated systems at very high stellar mass is driven by the fact that the simulation volume is only $\sim$100 Mpc and therefore very rare objects, like galaxies at the highest stellar masses, are absent from the simulation box.}
\label{fig:sfrate}
\end{figure}

\begin{center}
\begin{table*}
\centering
\begin{tabular}{|| c | c | c | c | c |}
\hline
\hline

1 & 2 & 3 & 4 & 5\\
\hline
Morphology & log$_{10}$ Sim. SFR [M$_{\odot}$yr$^{-1}$] & log$_{10}$ Obs. SFR [M$_{\odot}$ yr$^{-1}$]  & log$_{10}$ Sim. sSFR [yr$^{-1}$] & log$_{10}$ Obs. sSFR [yr$^{-1}$] \\
 & [16$^{\rm{th}}$ $\bullet$ \textbf{median} $\bullet$ 84$^{\rm{th}}$] & [16$^{\rm{th}}$ $\bullet$ \textbf{median} $\bullet$ 84$^{\rm{th}}$] & [16$^{\rm{th}}$ $\bullet$ \textbf{median} $\bullet$ 84$^{\rm{th}}$] & [16$^{\rm{th}}$ $\bullet$ \textbf{median} $\bullet$ 84$^{\rm{th}}$] \\
\hline
\hline
Discs & 0.28 $\bullet$ \textbf{0.76} $\bullet$ 1.07 & 0.23 $\bullet$ \textbf{0.65} $\bullet$ 0.86  & -11.13 $\bullet$ \textbf{-10.68} $\bullet$ -10.36 & -11.53 $\bullet$ \textbf{-10.94} $\bullet$ -10.67\\
\\
Spheroids & -0.73 $\bullet$ \textbf{0.12} $\bullet$ 0.68 & -0.94 $\bullet$ \textbf{-0.37} $\bullet$ 0.31 & -12.18 $\bullet$ \textbf{-11.00} $\bullet$ -10.39 & -12.56 $\bullet$ \textbf{-11.96} $\bullet$ -11.24 \\
\\
\end{tabular}
\caption{SFRs and sSFRs for massive (M$_*$ $>$ 10$^{11.4}$ M$_\odot$) galaxies of different morphologies (indicated in column 1). In each column we describe the 16th, 50th (i.e. the median) and 84th percentile values from the SFR and sSFR distributions. Columns are as follows: (2) total SFRs of simulated massive galaxies from J20 (3) total SFRs of the observed massive galaxies, calculated using SED-fitting of UV + optical photometry, from the GSWLC (4) total sSFRs of simulated massive galaxies from J20 (5) total sSFRs of the observed massive galaxies. The median values of the simulated and observed SFRs and sSFRs are in reasonably good agreement with each other for the massive disc population.}
\label{tab:sfr}
\end{table*}
\end{center}

In Figure \ref{fig:sfrate}, we compare the total SFRs of the observed massive discs calculated using integrated UV and optical photometry (blue) with the corresponding theoretical predictions (where the \textbf{total} SFR is calculated as an average over $\sim$100 Myrs) from J20 (red). The choice of 100 Myr is driven by the fact that GALEX UV photometry (which is used to derive the observed SFRs) is most sensitive over this timescale \citep[e.g.][]{Morrissey2007}, making the theoretical and observed SFRs comparable. Both the median values of the observed and theoretical samples (shown using the dashed lines) and the distributions of their SFRs are in good agreement with each other (the spheroids are omitted for clarity). In Table \ref{tab:sfr}, we summarise the theoretical and observed SFRs of our massive galaxies. For completeness, we also show the specific SFRs (sSFRs) of these systems. %finding that these are also consistent between the the observed galaxies and J20. %For completeness, we also show, in Table \ref{tab:sfr}, the spectroscopic SFRs of our observational sample, which are calculated using H$\alpha$ fluxes in the SDSS spectroscopic fibre. Not unexpectedly, the spectroscopic fibre SFRs are significantly lower than the total photometric SFRs, because they sample only the central regions of the galaxies. 
Finally, we note that, in a similar vein to what is seen in the predictions of J20, the SFRs and sSFRs of the observed massive discs are significantly elevated compared to that in their spheroidal counterparts.

\begin{figure}
\centering
\includegraphics[width=0.9\columnwidth]{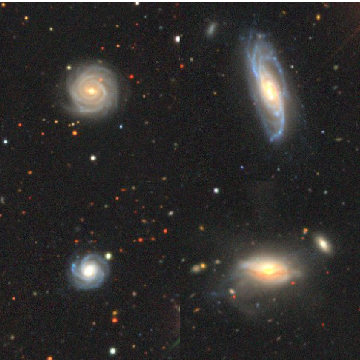}
\caption{DECaLS images of the 4 galaxies which have HI detections from the ALFALFA survey. All 4 systems are discs, with only 1 showing evidence of tidal features (lower right-hand panel). This disturbed disc has a high signal-to-noise detection in ALFALFA, while the other three have low S/N detections, due to their lower HI content (see Figure \ref{fig:gas_mass} below).}
\label{fig:gas_gals}
\end{figure}

\begin{figure*}
\centering
% \includegraphics[width=0.95\columnwidth]{figures/gas_mass.png}
% \hspace{0.5cm}
\includegraphics[width=0.95\textwidth]{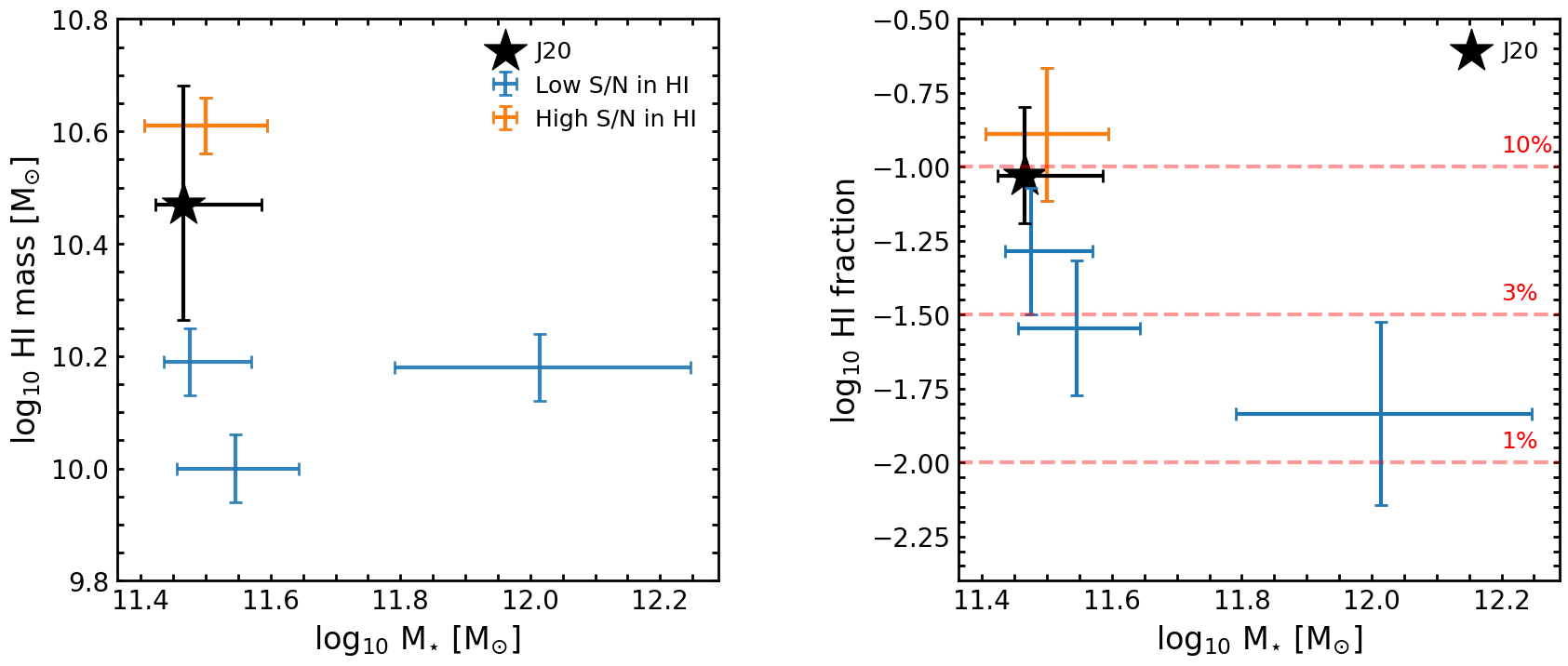}
\caption{\textbf{Left:} HI mass vs stellar mass for galaxies that have HI detections in the ALFALFA survey. The disturbed disc (lower right-hand panel in Figure \ref{fig:gas_gals}) has a high signal-to-noise (S/N) detection in ALFALFA (orange) while the other three have low S/N detections (blue). \textbf{Right:} HI fraction (defined as HI mass divided by the sum of the HI and stellar masses) vs stellar mass for the galaxies in the left-hand panel. The median values and distributions for the HI gas masses and gas fractions from the simulated galaxy sample of J20 are shown in black. The HI mass, and the corresponding HI fraction, in the disturbed disc is higher than those in its relaxed counterparts, consistent with the idea that the gas is likely brought in by a merger and is then used up to form stars as this merger-driven starburst progresses and the tidal features fade away, as suggested by J20.}
\label{fig:gas_mass}
\end{figure*}

While the agreement between the observed and theoretical SFRs indicates that the gas masses involved in the star formation events are broadly predicted correctly, we explore empirical constraints on what these gas masses are likely to be. While large area surveys of molecular gas are not available, we explore the HI masses of our massive galaxies using the ALFALFA survey. None of the massive spheroids are detected in ALFALFA. However, 4 massive discs (out of 92) have ALFALFA detections. In all cases, both the spatial and velocity offsets between the SDSS and the ALFALFA sources are small, indicating reliable matches. The spatial offsets are all less than 15 arcseconds (the average positional accuracy of ALFALFA is $\sim$24 arcseconds), while the velocity offsets (calculated from the spectroscopic redshifts and HI velocities) are within 80 km s$^{-1}$. 

Figure \ref{fig:gas_gals} shows DECaLS images of these discs, three of which are relaxed and one is disturbed (lower-right hand panel). Figure \ref{fig:gas_mass} indicates that the HI mass (and corresponding HI fraction, defined as the HI mass divided by the sum of the HI and stellar masses) in the disturbed disc is higher than those in its relaxed counterparts. Note that the average mass of the observed massive discs is $10^{11.56}$ M$_{\odot}$ so the four galaxies in Figure \ref{fig:gas_gals} are not anomalous in mass compared to the rest of the massive disc population. The higher HI mass seen in the disturbed system is consistent with the idea that the gas is likely brought in by a merger and is then used up to form stars, as this merger-driven starburst progresses and the tidal features fade away. The range of HI masses and fractions in massive galaxies from J20 are shown for comparison and are consistent with that in the disturbed disc.  

It is interesting to explore the possible gas fractions of the satellites that trigger the formation of these massive disc galaxies. To do this, we consider the HI content of the disturbed disc ($\sim$ 10$^{10.6}$ M$_{\odot}$). If the hypothesis in J20 is correct, then this is likely to be more representative of the original gas mass brought in by the minor merger, since, in the relaxed systems, much of the gas is likely to have been used up by the merger-driven star formation episode before we observe the system. Using the median merger mass ratio that creates the massive discs from J20 ($\sim$1:4.3) implies that the HI fraction of the satellite could be between $\sim$39 per cent (if one assumes the lower limit of the HI mass and the upper limit of the stellar mass) and $\sim$76 per cent (if one assumes the upper limit of the HI mass and the lower limit of the stellar mass). 

While we caution that these numbers are only indicative they do suggest that the satellites are indeed likely to be gas-rich, consistent with the theoretical predictions. It is worth noting that the satellites themselves have to be relatively massive galaxies. For example, assuming a merger mass ratio of $\sim$1:4.3 and a larger galaxy with a stellar mass of 10$^{11.4}$ M$_{\odot}$, the satellite would have a stellar mass of $\sim$ 10$^{10.8}$ M$_{\odot}$). The lower end of the estimated gas fractions ($\sim$39 per cent) is a factor of $\sim$3 higher than the median gas fractions for such systems at low redshift \citep[e.g.][]{Zhang2009,Catinella2010,Calette2018,Hunt2020}. An important caveat in this analysis of HI properties is that it is based on a very small sample of objects. Dedicated deep HI imaging, or future HI surveys which are deeper than those currently available, are required to provide true statistical insights into the empirical gas properties of the massive disc population and consolidate the evidence from the optical images of their minor-merger-driven origin.

%.....................................................................

\section{Summary}
\label{sec:summary}

In our standard structure-formation paradigm, the morphological transformation of massive galaxies, from discs to spheroids, is thought to be driven largely by merging. Furthermore, the merger activity experienced by galaxies tends to be a strong function of stellar mass, with the most massive galaxies having the richest merger histories. It is, therefore, surprising that a significant minority (>10 per cent) of massive galaxies at the highest stellar masses (M$_{\star}>10^{11.4}$ M$_{\odot}$), are in fact discs. In J20, we have used a cosmological hydrodynamical simulation to show that extremely massive discs can form via one of two channels. The dominant channel is a minor merger between a spheroid and a gas-rich satellite, while the secondary channel involves a disc galaxy maintaining its discy morphology due to an anomalously quiet merger history. In this paper, we have studied a large statistical sample of nearby massive galaxies, using data from the SDSS, GALEX, DECaLS and ALFALFA surveys to explore whether the observations support the predicted properties and formation mechanisms of massive discs in J20. Our main conclusions are as follows:

\begin{itemize}
    \item Massive discs account for $\sim$13 per cent of massive galaxies in our observational sample, in good agreement with the predicted fraction of discs in J20 ($\sim$11 per cent).
    
    \item $\sim$64 per cent of our massive discs show evidence for tidal features (compared to $\sim$31 per cent of our massive spheroids). This is a lower limit because deeper images are likely to reveal more galaxies with tidal features (see Figure \ref{fig:depth_gals}). In contrast, the tidal feature fraction in low-mass discs, in images that have similar depth and seeing, is $\sim$11-17 per cent, depending on the specific morphology of the low-mass disc population in question. The presence of tidal features is therefore significantly elevated in massive discs, by at least a factor of 4 compared to their low-mass counterparts, indicating a significant role for merging in the formation of these systems. 
    
    \item The major-merger fraction for massive galaxies at low redshift is only a few per cent, with tidal features remaining visible for 2-3 Gyrs at the depth of the DECaLS images. The majority of the interactions seen in our massive discs are, therefore, likely to be minor mergers (i.e. those with mass ratios less than $\sim$1:4), which are several times more frequent than their major counterparts. This is consistent with the prediction in J20 that minor mergers are the principal formation mechanism for massive disc galaxies. 
    
    \item The SFRs of our observed massive discs are in good agreement with those in their simulated counterparts in J20 and also show the predicted elevation compared to that in the massive spheroids. This suggests that the minor mergers in the massive discs are indeed gas-rich, consistent with the hypothesis presented in J20.
    
    \item While none of the massive spheroids are detected in HI, four massive discs have HI detections, of which three are relaxed and one is disturbed. The HI mass (and corresponding HI fraction) in the disturbed disc is higher than those in its relaxed counterparts, consistent with the idea that the gas is brought in by a merger, and is then used up to form stars as this merger-driven starburst progresses and the tidal features fade away. 
    
    \item Combining the HI content of the disturbed disc ($\sim$10$^{10.6}$ M$_{\odot}$), which is likely to be representative of the original gas mass brought in by this particular merger, and the median merger mass ratio predicted in J20 for the massive discs, suggests that the gas fractions of the accreted satellites are likely to be extremely high ($\sim$40 per cent or more). Since the satellites themselves are relatively massive galaxies (e.g. assuming that the merger has a mass ratio of $\sim$1:4 and the larger galaxy has a stellar mass of 10$^{11.4}$ M$_{\odot}$, M$_{\rm{sat}}$ is 10$^{10.8}$ M$_{\odot}$), such gas fractions are a factor of $\sim$3 higher than the median gas fractions for such systems at low redshift. This is consistent with the predictions of J20 that massive discs are formed in gas-rich minor mergers. 
\end{itemize}

In summary, the observed properties of nearby massive discs, in terms of their morphological fractions, SFRs and HI properties are in good agreement with the theoretical predictions of J20. This, in turn, suggests that massive disc galaxies in the nearby Universe are likely to have been formed primarily via minor mergers between spheroids and gas-rich satellites. 

%.....................................................................

\section*{Acknowledgements}

We thank the anonymous referee for many constructive suggestions that improved the quality of the original manuscript. We thank Timothy Davis for many useful discussions. This work was supported in part by the Yonsei University Research Fund (Yonsei Frontier Lab, Young Researcher Supporting Program) of 2021 and by the Korean National Research Foundation (NRF-2020R1A2C3003769). RAJ and SK acknowledge support from the STFC [ST/R504786/1, ST/S00615X/1]. SK acknowledges a Senior Research Fellowship from Worcester College Oxford.

% \textbf{Add acknowledgements for GALEX and ALFALFA here.}\\
GALEX (Galaxy Evolution Explorer) is a NASA Small Explorer, launched in 2003 April. We gratefully acknowledge NASA’s support for construction, operation, and science analysis for the GALEX mission, developed in cooperation with the Centre National d’Etudes Spatiales of France and the Korean Ministry of Science and Technology.

We acknowledge the work of the ALFALFA team in observing, flagging, and processing the ALFALFA data that this work makes use of. The ALFALFA team at Cornell is supported by NSF grants AST-0607007 and AST-1107390 and by the Brinson Foundation.

Funding for the SDSS and SDSS-II has been provided by the Alfred P. Sloan Foundation, the Participating Institutions, the National Science Foundation, the U.S. Department of Energy, the National Aeronautics and Space Administration, the Japanese Monbukagakusho, the Max Planck Society, and the Higher Education Funding Council for England. The SDSS Web Site is http://www.sdss.org/.

The SDSS is managed by the Astrophysical Research Consortium for the Participating Institutions. The Participating Institutions are the American Museum of Natural History, Astrophysical Institute Potsdam, University of Basel, University of Cambridge, Case Western Reserve University, University of Chicago, Drexel University, Fermilab, the Institute for Advanced Study, the Japan Participation Group, Johns Hopkins University, the Joint Institute for Nuclear Astrophysics, the Kavli Institute for Particle Astrophysics and Cosmology, the Korean Scientist Group, the Chinese Academy of Sciences (LAMOST), Los Alamos National Laboratory, the Max-Planck-Institute for Astronomy (MPIA), the Max-Planck-Institute for Astrophysics (MPA), New Mexico State University, Ohio State University, University of Pittsburgh, University of Portsmouth, Princeton University, the United States Naval Observatory, and the University of Washington.

The Legacy Surveys consist of three individual and complementary projects: the Dark Energy Camera Legacy Survey (DECaLS; Proposal ID \#2014B-0404; PIs: David Schlegel and Arjun Dey), the Beijing-Arizona Sky Survey (BASS; NOAO Prop. ID \#2015A-0801; PIs: Zhou Xu and Xiaohui Fan), and the Mayall z-band Legacy Survey (MzLS; Prop. ID \#2016A-0453; PI: Arjun Dey). DECaLS, BASS and MzLS together include data obtained, respectively, at the Blanco telescope, Cerro Tololo Inter-American Observatory, NSF’s NOIRLab; the Bok telescope, Steward Observatory, University of Arizona; and the Mayall telescope, Kitt Peak National Observatory, NOIRLab. The Legacy Surveys project is honored to be permitted to conduct astronomical research on Iolkam Du’ag (Kitt Peak), a mountain with particular significance to the Tohono O’odham Nation.

NOIRLab is operated by the Association of Universities for Research in Astronomy (AURA) under a cooperative agreement with the National Science Foundation.

This project used data obtained with the Dark Energy Camera (DECam), which was constructed by the Dark Energy Survey (DES) collaboration. Funding for the DES Projects has been provided by the U.S. Department of Energy, the U.S. National Science Foundation, the Ministry of Science and Education of Spain, the Science and Technology Facilities Council of the United Kingdom, the Higher Education Funding Council for England, the National Center for Supercomputing Applications at the University of Illinois at Urbana-Champaign, the Kavli Institute of Cosmological Physics at the University of Chicago, Center for Cosmology and Astro-Particle Physics at the Ohio State University, the Mitchell Institute for Fundamental Physics and Astronomy at Texas A\&M University, Financiadora de Estudos e Projetos, Fundacao Carlos Chagas Filho de Amparo, Financiadora de Estudos e Projetos, Fundacao Carlos Chagas Filho de Amparo a Pesquisa do Estado do Rio de Janeiro, Conselho Nacional de Desenvolvimento Cientifico e Tecnologico and the Ministerio da Ciencia, Tecnologia e Inovacao, the Deutsche Forschungsgemeinschaft and the Collaborating Institutions in the Dark Energy Survey. The Collaborating Institutions are Argonne National Laboratory, the University of California at Santa Cruz, the University of Cambridge, Centro de Investigaciones Energeticas, Medioambientales y Tecnologicas-Madrid, the University of Chicago, University College London, the DES-Brazil Consortium, the University of Edinburgh, the Eidgenossische Technische Hochschule (ETH) Zurich, Fermi National Accelerator Laboratory, the University of Illinois at Urbana-Champaign, the Institut de Ciencies de l’Espai (IEEC/CSIC), the Institut de Fisica d’Altes Energies, Lawrence Berkeley National Laboratory, the Ludwig Maximilians Universitat Munchen and the associated Excellence Cluster Universe, the University of Michigan, NSF’s NOIRLab, the University of Nottingham, the Ohio State University, the University of Pennsylvania, the University of Portsmouth, SLAC National Accelerator Laboratory, Stanford University, the University of Sussex, and Texas A\&M University.

BASS is a key project of the Telescope Access Program (TAP), which has been funded by the National Astronomical Observatories of China, the Chinese Academy of Sciences (the Strategic Priority Research Program “The Emergence of Cosmological Structures” Grant \#XDB09000000), and the Special Fund for Astronomy from the Ministry of Finance. The BASS is also supported by the External Cooperation Program of Chinese Academy of Sciences (Grant \#114A11KYSB20160057), and Chinese National Natural Science Foundation (Grant \#11433005).

The Legacy Survey team makes use of data products from the Near-Earth Object Wide-field Infrared Survey Explorer (NEOWISE), which is a project of the Jet Propulsion Laboratory/California Institute of Technology. NEOWISE is funded by the National Aeronautics and Space Administration.

The Legacy Surveys imaging of the DESI footprint is supported by the Director, Office of Science, Office of High Energy Physics of the U.S. Department of Energy under Contract No. DE-AC02-05CH1123, by the National Energy Research Scientific Computing Center, a DOE Office of Science User Facility under the same contract; and by the U.S. National Science Foundation, Division of Astronomical Sciences under Contract No. AST-0950945 to NOAO.

%.....................................................................

\section*{Data Availability}

Please contact the authors if the sample of observed massive galaxies used in this study is of interest.

%........................................................

\bibliographystyle{mnras}
\bibliography{bib}

\begin{thebibliography}{}
\makeatletter
\relax
\def\mn@urlcharsother{\let\do\@makeother \do\$\do\&\do\#\do\^\do\_\do\%\do\~}
\def\mn@doi{\begingroup\mn@urlcharsother \@ifnextchar [ {\mn@doi@}
  {\mn@doi@[]}}
\def\mn@doi@[#1]#2{\def\@tempa{#1}\ifx\@tempa\@empty \href
  {http://dx.doi.org/#2} {doi:#2}\else \href {http://dx.doi.org/#2} {#1}\fi
  \endgroup}
\def\mn@eprint#1#2{\mn@eprint@#1:#2::\@nil}
\def\mn@eprint@arXiv#1{\href {http://arxiv.org/abs/#1} {{\tt arXiv:#1}}}
\def\mn@eprint@dblp#1{\href {http://dblp.uni-trier.de/rec/bibtex/#1.xml}
  {dblp:#1}}
\def\mn@eprint@#1:#2:#3:#4\@nil{\def\@tempa {#1}\def\@tempb {#2}\def\@tempc
  {#3}\ifx \@tempc \@empty \let \@tempc \@tempb \let \@tempb \@tempa \fi \ifx
  \@tempb \@empty \def\@tempb {arXiv}\fi \@ifundefined
  {mn@eprint@\@tempb}{\@tempb:\@tempc}{\expandafter \expandafter \csname
  mn@eprint@\@tempb\endcsname \expandafter{\@tempc}}}

\bibitem[\protect\citeauthoryear{{Abazajian} et~al.,}{{Abazajian}
  et~al.}{2009}]{Abazajian2009}
{Abazajian} K.~N.,  et~al., 2009, \mn@doi [\apjs]
  {10.1088/0067-0049/182/2/543}, \href
  {https://ui.adsabs.harvard.edu/abs/2009ApJS..182..543A} {182, 543}

\bibitem[\protect\citeauthoryear{{Aihara} et~al.,}{{Aihara}
  et~al.}{2019}]{Aihara2019}
{Aihara} H.,  et~al., 2019, \mn@doi [\pasj] {10.1093/pasj/psz103}, \href
  {https://ui.adsabs.harvard.edu/abs/2019PASJ..tmp..106A} {p.~106}

\bibitem[\protect\citeauthoryear{{Barnes}}{{Barnes}}{1992}]{Barnes1992}
{Barnes} J.~E.,  1992, \mn@doi [\apj] {10.1086/171522}, \href
  {http://adsabs.harvard.edu/abs/1992ApJ...393..484B} {393, 484}

\bibitem[\protect\citeauthoryear{{Bernardi} et~al.,}{{Bernardi}
  et~al.}{2003}]{Bernardi2003}
{Bernardi} M.,  et~al., 2003, \mn@doi [\aj] {10.1086/367776}, \href
  {http://adsabs.harvard.edu/abs/2003AJ....125.1817B} {125, 1817}

\bibitem[\protect\citeauthoryear{{Blumenthal} et~al.,}{{Blumenthal}
  et~al.}{2020}]{Blumenthal2020}
{Blumenthal} K.~A.,  et~al., 2020, \mn@doi [\mnras] {10.1093/mnras/stz3472},
  \href {https://ui.adsabs.harvard.edu/abs/2020MNRAS.492.2075B} {492, 2075}

\bibitem[\protect\citeauthoryear{{Bournaud}, {Jog}  \& {Combes}}{{Bournaud}
  et~al.}{2007}]{Bournaud2007}
{Bournaud} F.,  {Jog} C.~J.,   {Combes} F.,  2007, \mn@doi [\aap]
  {10.1051/0004-6361:20078010}, \href
  {http://adsabs.harvard.edu/abs/2007A%26A...476.1179B} {476, 1179}

\bibitem[\protect\citeauthoryear{{Brinchmann}, {Charlot}, {White}, {Tremonti},
  {Kauffmann}, {Heckman}  \& {Brinkmann}}{{Brinchmann}
  et~al.}{2004}]{Brinchmann2004}
{Brinchmann} J.,  {Charlot} S.,  {White} S.~D.~M.,  {Tremonti} C.,  {Kauffmann}
  G.,  {Heckman} T.,   {Brinkmann} J.,  2004, \mn@doi [\mnras]
  {10.1111/j.1365-2966.2004.07881.x}, \href
  {https://ui.adsabs.harvard.edu/abs/2004MNRAS.351.1151B} {351, 1151}

\bibitem[\protect\citeauthoryear{{Buitrago}, {Conselice}, {Epinat}, {Bedregal},
  {Gr{\"u}tzbauch}  \& {Weiner}}{{Buitrago} et~al.}{2014}]{Buitrago2014}
{Buitrago} F.,  {Conselice} C.~J.,  {Epinat} B.,  {Bedregal} A.~G.,
  {Gr{\"u}tzbauch} R.,   {Weiner} B.~J.,  2014, \mn@doi [\mnras]
  {10.1093/mnras/stu034}, \href
  {http://adsabs.harvard.edu/abs/2014MNRAS.439.1494B} {439, 1494}

\bibitem[\protect\citeauthoryear{{Calette}, {Avila-Reese},
  {Rodr{\'\i}guez-Puebla}, {Hern{\'a}ndez-Toledo}  \& {Papastergis}}{{Calette}
  et~al.}{2018}]{Calette2018}
{Calette} A.~R.,  {Avila-Reese} V.,  {Rodr{\'\i}guez-Puebla} A.,
  {Hern{\'a}ndez-Toledo} H.,   {Papastergis} E.,  2018, \rmxaa, \href
  {https://ui.adsabs.harvard.edu/abs/2018RMxAA..54..443C} {54, 443}

\bibitem[\protect\citeauthoryear{{Catinella} et~al.,}{{Catinella}
  et~al.}{2010}]{Catinella2010}
{Catinella} B.,  et~al., 2010, \mn@doi [\mnras]
  {10.1111/j.1365-2966.2009.16180.x}, \href
  {https://ui.adsabs.harvard.edu/abs/2010MNRAS.403..683C} {403, 683}

\bibitem[\protect\citeauthoryear{{Cattaneo}, {Mamon}, {Warnick}  \&
  {Knebe}}{{Cattaneo} et~al.}{2011}]{Cattaneo2011}
{Cattaneo} A.,  {Mamon} G.~A.,  {Warnick} K.,   {Knebe} A.,  2011, \mn@doi
  [\aap] {10.1051/0004-6361/201015780}, \href
  {https://ui.adsabs.harvard.edu/abs/2011A&A...533A...5C} {533, A5}

\bibitem[\protect\citeauthoryear{{Conselice}}{{Conselice}}{2006}]{Conselice2006}
{Conselice} C.~J.,  2006, \mn@doi [\apj] {10.1086/499067}, \href
  {http://adsabs.harvard.edu/abs/2006ApJ...638..686C} {638, 686}

\bibitem[\protect\citeauthoryear{{Conselice}, {Bluck}, {Mortlock}, {Palamara}
  \& {Benson}}{{Conselice} et~al.}{2014}]{Conselice2014}
{Conselice} C.~J.,  {Bluck} A.~F.~L.,  {Mortlock} A.,  {Palamara} D.,
  {Benson} A.~J.,  2014, \mn@doi [\mnras] {10.1093/mnras/stu1385}, \href
  {http://adsabs.harvard.edu/abs/2014MNRAS.444.1125C} {444, 1125}

\bibitem[\protect\citeauthoryear{{Darg} et~al.,}{{Darg}
  et~al.}{2010}]{Darg2010}
{Darg} D.~W.,  et~al., 2010, \mn@doi [\mnras]
  {10.1111/j.1365-2966.2009.15686.x}, \href
  {https://ui.adsabs.harvard.edu/abs/2010MNRAS.401.1043D} {401, 1043}

\bibitem[\protect\citeauthoryear{{Dey} et~al.,}{{Dey} et~al.}{2019}]{Dey2019}
{Dey} A.,  et~al., 2019, \mn@doi [\aj] {10.3847/1538-3881/ab089d}, \href
  {https://ui.adsabs.harvard.edu/abs/2019AJ....157..168D} {157, 168}

\bibitem[\protect\citeauthoryear{{Di Matteo}, {Combes}, {Melchior}  \&
  {Semelin}}{{Di Matteo} et~al.}{2007}]{DiMatteo2007}
{Di Matteo} P.,  {Combes} F.,  {Melchior} A.-L.,   {Semelin} B.,  2007, \mn@doi
  [\aap] {10.1051/0004-6361:20066959}, \href
  {http://adsabs.harvard.edu/abs/2007A%26A...468...61D} {468, 61}

\bibitem[\protect\citeauthoryear{{Dubois}, {Gavazzi}, {Peirani}  \&
  {Silk}}{{Dubois} et~al.}{2013}]{Dubois2013}
{Dubois} Y.,  {Gavazzi} R.,  {Peirani} S.,   {Silk} J.,  2013, \mn@doi [\mnras]
  {10.1093/mnras/stt997}, \href
  {http://adsabs.harvard.edu/abs/2013MNRAS.433.3297D} {433, 3297}

\bibitem[\protect\citeauthoryear{{Dubois} et~al.,}{{Dubois}
  et~al.}{2014}]{Dubois2014}
{Dubois} Y.,  et~al., 2014, \mn@doi [\mnras] {10.1093/mnras/stu1227}, \href
  {http://adsabs.harvard.edu/abs/2014MNRAS.444.1453D} {444, 1453}

\bibitem[\protect\citeauthoryear{{Dubois}, {Peirani}, {Pichon}, {Devriendt},
  {Gavazzi}, {Welker}  \& {Volonteri}}{{Dubois} et~al.}{2016}]{Dubois2016}
{Dubois} Y.,  {Peirani} S.,  {Pichon} C.,  {Devriendt} J.,  {Gavazzi} R.,
  {Welker} C.,   {Volonteri} M.,  2016, \mn@doi [\mnras]
  {10.1093/mnras/stw2265}, \href
  {http://adsabs.harvard.edu/abs/2016MNRAS.463.3948D} {463, 3948}

\bibitem[\protect\citeauthoryear{{Faber} et~al.,}{{Faber}
  et~al.}{2007}]{Faber2007}
{Faber} S.~M.,  et~al., 2007, \mn@doi [\apj] {10.1086/519294}, \href
  {http://adsabs.harvard.edu/abs/2007ApJ...665..265F} {665, 265}

\bibitem[\protect\citeauthoryear{{Font}, {McCarthy}, {Le Brun}, {Crain}  \&
  {Kelvin}}{{Font} et~al.}{2017}]{Font2017}
{Font} A.~S.,  {McCarthy} I.~G.,  {Le Brun} A.~M.~C.,  {Crain} R.~A.,
  {Kelvin} L.~S.,  2017, preprint, \href
  {http://adsabs.harvard.edu/abs/2017arXiv171000415F} {} (\mn@eprint {arXiv}
  {1710.00415})

\bibitem[\protect\citeauthoryear{{Governato} et~al.,}{{Governato}
  et~al.}{2009}]{Governato2009}
{Governato} F.,  et~al., 2009, \mn@doi [\mnras]
  {10.1111/j.1365-2966.2009.15143.x}, \href
  {https://ui.adsabs.harvard.edu/abs/2009MNRAS.398..312G} {398, 312}

\bibitem[\protect\citeauthoryear{{Haynes} et~al.,}{{Haynes}
  et~al.}{2011}]{Haynes2011}
{Haynes} M.~P.,  et~al., 2011, \mn@doi [\aj] {10.1088/0004-6256/142/5/170},
  \href {https://ui.adsabs.harvard.edu/abs/2011AJ....142..170H} {142, 170}

\bibitem[\protect\citeauthoryear{{Hilz}, {Naab}  \& {Ostriker}}{{Hilz}
  et~al.}{2013}]{Hilz2013}
{Hilz} M.,  {Naab} T.,   {Ostriker} J.~P.,  2013, \mn@doi [\mnras]
  {10.1093/mnras/sts501}, \href
  {https://ui.adsabs.harvard.edu/abs/2013MNRAS.429.2924H} {429, 2924}

\bibitem[\protect\citeauthoryear{{Hood}, {Kannappan}, {Stark}, {Dell'Antonio},
  {Moffett}, {Eckert}, {Norris}  \& {Hendel}}{{Hood} et~al.}{2018}]{Hood2018}
{Hood} C.~E.,  {Kannappan} S.~J.,  {Stark} D.~V.,  {Dell'Antonio} I.~P.,
  {Moffett} A.~J.,  {Eckert} K.~D.,  {Norris} M.~A.,   {Hendel} D.,  2018,
  \mn@doi [\apj] {10.3847/1538-4357/aab719}, \href
  {https://ui.adsabs.harvard.edu/abs/2018ApJ...857..144H} {857, 144}

\bibitem[\protect\citeauthoryear{{Hopkins}, {Cox}, {Younger}  \&
  {Hernquist}}{{Hopkins} et~al.}{2009}]{Hopkins2009}
{Hopkins} P.~F.,  {Cox} T.~J.,  {Younger} J.~D.,   {Hernquist} L.,  2009,
  \mn@doi [\apj] {10.1088/0004-637X/691/2/1168}, \href
  {https://ui.adsabs.harvard.edu/abs/2009ApJ...691.1168H} {691, 1168}

\bibitem[\protect\citeauthoryear{{Hunt}, {Tortora}, {Ginolfi}  \&
  {Schneider}}{{Hunt} et~al.}{2020}]{Hunt2020}
{Hunt} L.~K.,  {Tortora} C.,  {Ginolfi} M.,   {Schneider} R.,  2020, \mn@doi
  [\aap] {10.1051/0004-6361/202039021}, \href
  {https://ui.adsabs.harvard.edu/abs/2020A&A...643A.180H} {643, A180}

\bibitem[\protect\citeauthoryear{{Jackson}, {Martin}, {Kaviraj}, {Laigle},
  {Devriendt}, {Dubois}  \& {Pichon}}{{Jackson} et~al.}{2020}]{Jackson2020disc}
{Jackson} R.~A.,  {Martin} G.,  {Kaviraj} S.,  {Laigle} C.,  {Devriendt}
  J.~E.~G.,  {Dubois} Y.,   {Pichon} C.,  2020, \mn@doi [\mnras]
  {10.1093/mnras/staa970}, \href
  {https://ui.adsabs.harvard.edu/abs/2020MNRAS.494.5568J} {494, 5568}

\bibitem[\protect\citeauthoryear{{Jackson} et~al.,}{{Jackson}
  et~al.}{2021}]{Jackson2021}
{Jackson} R.~A.,  et~al., 2021, \mn@doi [\mnras] {10.1093/mnras/stab093}, \href
  {https://ui.adsabs.harvard.edu/abs/2021MNRAS.502.1785J} {502, 1785}

\bibitem[\protect\citeauthoryear{{Jogee} et~al.,}{{Jogee}
  et~al.}{2009}]{Jogee2009}
{Jogee} S.,  et~al., 2009, \mn@doi [\apj] {10.1088/0004-637X/697/2/1971}, \href
  {https://ui.adsabs.harvard.edu/abs/2009ApJ...697.1971J} {697, 1971}

\bibitem[\protect\citeauthoryear{{Kaviraj}}{{Kaviraj}}{2010}]{Kaviraj2010}
{Kaviraj} S.,  2010, \mn@doi [\mnras] {10.1111/j.1365-2966.2010.16714.x}, \href
  {http://adsabs.harvard.edu/abs/2010MNRAS.406..382K} {406, 382}

\bibitem[\protect\citeauthoryear{{Kaviraj}}{{Kaviraj}}{2014a}]{Kaviraj2014a}
{Kaviraj} S.,  2014a, \mn@doi [\mnras] {10.1093/mnrasl/slt136}, \href
  {http://adsabs.harvard.edu/abs/2014MNRAS.437L..41K} {437, L41}

\bibitem[\protect\citeauthoryear{{Kaviraj}}{{Kaviraj}}{2014b}]{Kaviraj2014b}
{Kaviraj} S.,  2014b, \mn@doi [\mnras] {10.1093/mnras/stu338}, \href
  {http://adsabs.harvard.edu/abs/2014MNRAS.440.2944K} {440, 2944}

\bibitem[\protect\citeauthoryear{{Kaviraj}, {Tan}, {Ellis}  \&
  {Silk}}{{Kaviraj} et~al.}{2011}]{Kaviraj2011}
{Kaviraj} S.,  {Tan} K.-M.,  {Ellis} R.~S.,   {Silk} J.,  2011, \mn@doi
  [\mnras] {10.1111/j.1365-2966.2010.17754.x}, \href
  {http://adsabs.harvard.edu/abs/2011MNRAS.411.2148K} {411, 2148}

\bibitem[\protect\citeauthoryear{{Kaviraj}, {Devriendt}, {Dubois}, {Slyz},
  {Welker}, {Pichon}, {Peirani}  \& {Le Borgne}}{{Kaviraj}
  et~al.}{2015}]{Kaviraj2015}
{Kaviraj} S.,  {Devriendt} J.,  {Dubois} Y.,  {Slyz} A.,  {Welker} C.,
  {Pichon} C.,  {Peirani} S.,   {Le Borgne} D.,  2015, \mn@doi [\mnras]
  {10.1093/mnras/stv1500}, \href
  {http://adsabs.harvard.edu/abs/2015MNRAS.452.2845K} {452, 2845}

\bibitem[\protect\citeauthoryear{{Kaviraj} et~al.,}{{Kaviraj}
  et~al.}{2017}]{Kaviraj2017}
{Kaviraj} S.,  et~al., 2017, \mn@doi [\mnras] {10.1093/mnras/stx126}, \href
  {http://adsabs.harvard.edu/abs/2017MNRAS.467.4739K} {467, 4739}

\bibitem[\protect\citeauthoryear{{Kaviraj}, {Martin}  \& {Silk}}{{Kaviraj}
  et~al.}{2019}]{Kaviraj2019}
{Kaviraj} S.,  {Martin} G.,   {Silk} J.,  2019, \mn@doi [\mnras]
  {10.1093/mnrasl/slz102}, \href
  {https://ui.adsabs.harvard.edu/abs/2019MNRAS.489L..12K} {489, L12}

\bibitem[\protect\citeauthoryear{{Li} \& {White}}{{Li} \&
  {White}}{2009}]{Li2009}
{Li} C.,  {White} S. D.~M.,  2009, \mn@doi [\mnras]
  {10.1111/j.1365-2966.2009.15268.x}, \href
  {https://ui.adsabs.harvard.edu/abs/2009MNRAS.398.2177L} {398, 2177}

\bibitem[\protect\citeauthoryear{{Lofthouse}, {Kaviraj}, {Conselice},
  {Mortlock}  \& {Hartley}}{{Lofthouse} et~al.}{2017}]{Lofthouse2017}
{Lofthouse} E.~K.,  {Kaviraj} S.,  {Conselice} C.~J.,  {Mortlock} A.,
  {Hartley} W.,  2017, \mn@doi [\mnras] {10.1093/mnras/stw2895}, \href
  {http://adsabs.harvard.edu/abs/2017MNRAS.465.2895L} {465, 2895}

\bibitem[\protect\citeauthoryear{{L{\'o}pez-Sanjuan}, {Balcells},
  {P{\'e}rez-Gonz{\'a}lez}, {Barro}, {Garc{\'\i}a-Dab{\'o}}, {Gallego}  \&
  {Zamorano}}{{L{\'o}pez-Sanjuan} et~al.}{2010}]{LopezSanjuan2010}
{L{\'o}pez-Sanjuan} C.,  {Balcells} M.,  {P{\'e}rez-Gonz{\'a}lez} P.~G.,
  {Barro} G.,  {Garc{\'\i}a-Dab{\'o}} C.~E.,  {Gallego} J.,   {Zamorano} J.,
  2010, \mn@doi [\apj] {10.1088/0004-637X/710/2/1170}, \href
  {https://ui.adsabs.harvard.edu/abs/2010ApJ...710.1170L} {710, 1170}

\bibitem[\protect\citeauthoryear{{Lotz}, {Jonsson}, {Cox}, {Croton}, {Primack},
  {Somerville}  \& {Stewart}}{{Lotz} et~al.}{2011}]{Lotz2011}
{Lotz} J.~M.,  {Jonsson} P.,  {Cox} T.~J.,  {Croton} D.,  {Primack} J.~R.,
  {Somerville} R.~S.,   {Stewart} K.,  2011, \mn@doi [\apj]
  {10.1088/0004-637X/742/2/103}, \href
  {https://ui.adsabs.harvard.edu/abs/2011ApJ...742..103L} {742, 103}

\bibitem[\protect\citeauthoryear{{Mancillas}, {Duc}, {Combes}, {Bournaud},
  {Emsellem}, {Martig}  \& {Michel-Dansac}}{{Mancillas}
  et~al.}{2019}]{Mancillas2019}
{Mancillas} B.,  {Duc} P.-A.,  {Combes} F.,  {Bournaud} F.,  {Emsellem} E.,
  {Martig} M.,   {Michel-Dansac} L.,  2019, \mn@doi [\aap]
  {10.1051/0004-6361/201936320}, \href
  {https://ui.adsabs.harvard.edu/abs/2019A&A...632A.122M} {632, A122}

\bibitem[\protect\citeauthoryear{{Martig} et~al.,}{{Martig}
  et~al.}{2021}]{Martig2021}
{Martig} M.,  et~al., 2021, \mn@doi [\mnras] {10.1093/mnras/stab2729}, \href
  {https://ui.adsabs.harvard.edu/abs/2021MNRAS.508.2458M} {508, 2458}

\bibitem[\protect\citeauthoryear{{Martin}, {Kaviraj}, {Devriendt}, {Dubois}  \&
  {Pichon}}{{Martin} et~al.}{2018}]{Martin2018b}
{Martin} G.,  {Kaviraj} S.,  {Devriendt} J.~E.~G.,  {Dubois} Y.,   {Pichon} C.,
   2018, \mn@doi [\mnras] {10.1093/mnras/sty1936}, \href
  {http://adsabs.harvard.edu/abs/2018MNRAS.480.2266M} {480, 2266}

\bibitem[\protect\citeauthoryear{{Martin} et~al.,}{{Martin}
  et~al.}{2019}]{Martin2019}
{Martin} G.,  et~al., 2019, \mn@doi [\mnras] {10.1093/mnras/stz356}, \href
  {https://ui.adsabs.harvard.edu/abs/2019MNRAS.485..796M} {485, 796}

\bibitem[\protect\citeauthoryear{{Martin} et~al.,}{{Martin}
  et~al.}{2021}]{Martin2021}
{Martin} G.,  et~al., 2021, \mn@doi [\mnras] {10.1093/mnras/staa3443}, \href
  {https://ui.adsabs.harvard.edu/abs/2021MNRAS.500.4937M} {500, 4937}

\bibitem[\protect\citeauthoryear{{McIntosh}, {Guo}, {Hertzberg}, {Katz}, {Mo},
  {van den Bosch}  \& {Yang}}{{McIntosh} et~al.}{2008}]{McIntosh2008}
{McIntosh} D.~H.,  {Guo} Y.,  {Hertzberg} J.,  {Katz} N.,  {Mo} H.~J.,  {van
  den Bosch} F.~C.,   {Yang} X.,  2008, \mn@doi [\mnras]
  {10.1111/j.1365-2966.2008.13531.x}, \href
  {http://adsabs.harvard.edu/abs/2008MNRAS.388.1537M} {388, 1537}

\bibitem[\protect\citeauthoryear{{Monachesi}, {Bell}, {Radburn-Smith},
  {Bailin}, {de Jong}, {Holwerda}, {Streich}  \& {Silverstein}}{{Monachesi}
  et~al.}{2016}]{Monachesi2016}
{Monachesi} A.,  {Bell} E.~F.,  {Radburn-Smith} D.~J.,  {Bailin} J.,  {de Jong}
  R.~S.,  {Holwerda} B.,  {Streich} D.,   {Silverstein} G.,  2016, \mn@doi
  [\mnras] {10.1093/mnras/stv2987}, \href
  {https://ui.adsabs.harvard.edu/abs/2016MNRAS.457.1419M} {457, 1419}

\bibitem[\protect\citeauthoryear{{Morrissey} et~al.,}{{Morrissey}
  et~al.}{2007}]{Morrissey2007}
{Morrissey} P.,  et~al., 2007, \mn@doi [\apjs] {10.1086/520512}, \href
  {https://ui.adsabs.harvard.edu/abs/2007ApJS..173..682M} {173, 682}

\bibitem[\protect\citeauthoryear{{Mundy}, {Conselice}, {Duncan}, {Almaini},
  {H{\"a}u{\ss}ler}  \& {Hartley}}{{Mundy} et~al.}{2017}]{Mundy2017}
{Mundy} C.~J.,  {Conselice} C.~J.,  {Duncan} K.~J.,  {Almaini} O.,
  {H{\"a}u{\ss}ler} B.,   {Hartley} W.~G.,  2017, \mn@doi [\mnras]
  {10.1093/mnras/stx1238}, \href
  {https://ui.adsabs.harvard.edu/abs/2017MNRAS.470.3507M} {470, 3507}

\bibitem[\protect\citeauthoryear{{Nelson} et~al.,}{{Nelson}
  et~al.}{2019}]{Nelson2019}
{Nelson} D.,  et~al., 2019, \mn@doi [Computational Astrophysics and Cosmology]
  {10.1186/s40668-019-0028-x}, \href
  {https://ui.adsabs.harvard.edu/abs/2019ComAC...6....2N} {6, 2}

\bibitem[\protect\citeauthoryear{Ogle, Lanz, Nader  \& Helou}{Ogle
  et~al.}{2016}]{Ogle2016}
Ogle P.~M.,  Lanz L.,  Nader C.,   Helou G.,  2016, \mn@doi [The Astrophysical
  Journal] {10.3847/0004-637x/817/2/109}, 817, 109

\bibitem[\protect\citeauthoryear{Ogle, {Lanz}, {Appleton}, {Helou}  \&
  {Mazzarella}}{Ogle et~al.}{2019}]{Ogle2019}
Ogle P.~M.,  {Lanz} L.,  {Appleton} P.~N.,  {Helou} G.,   {Mazzarella} J.,
  2019, \mn@doi [\apjs] {10.3847/1538-4365/ab21c3}, \href
  {https://ui.adsabs.harvard.edu/abs/2019ApJS..243...14O} {243, 14}

\bibitem[\protect\citeauthoryear{{Oser}, {Ostriker}, {Naab}, {Johansson}  \&
  {Burkert}}{{Oser} et~al.}{2010}]{Oser2010}
{Oser} L.,  {Ostriker} J.~P.,  {Naab} T.,  {Johansson} P.~H.,   {Burkert} A.,
  2010, \mn@doi [\apj] {10.1088/0004-637X/725/2/2312}, \href
  {http://adsabs.harvard.edu/abs/2010ApJ...725.2312O} {725, 2312}

\bibitem[\protect\citeauthoryear{{Peschken}, {{\L}okas}  \&
  {Athanassoula}}{{Peschken} et~al.}{2019}]{Peschken2019}
{Peschken} N.,  {{\L}okas} E.~L.,   {Athanassoula} E.,  2019, arXiv e-prints,
  \href {https://ui.adsabs.harvard.edu/abs/2019arXiv190901033P} {p.
  arXiv:1909.01033}

\bibitem[\protect\citeauthoryear{{Robertson}, {Bullock}, {Cox}, {Di Matteo},
  {Hernquist}, {Springel}  \& {Yoshida}}{{Robertson}
  et~al.}{2006}]{Robertson2006}
{Robertson} B.,  {Bullock} J.~S.,  {Cox} T.~J.,  {Di Matteo} T.,  {Hernquist}
  L.,  {Springel} V.,   {Yoshida} N.,  2006, \mn@doi [\apj] {10.1086/504412},
  \href {http://adsabs.harvard.edu/abs/2006ApJ...645..986R} {645, 986}

\bibitem[\protect\citeauthoryear{{Rodriguez-Gomez} et~al.,}{{Rodriguez-Gomez}
  et~al.}{2015}]{RodriguezGomez2015}
{Rodriguez-Gomez} V.,  et~al., 2015, \mn@doi [\mnras] {10.1093/mnras/stv264},
  \href {https://ui.adsabs.harvard.edu/abs/2015MNRAS.449...49R} {449, 49}

\bibitem[\protect\citeauthoryear{{Ryan} R.~E. et~al.,}{{Ryan}
  et~al.}{2012}]{Ryan2012}
{Ryan} R.~E. J.,  et~al., 2012, \mn@doi [\apj] {10.1088/0004-637X/749/1/53},
  \href {https://ui.adsabs.harvard.edu/abs/2012ApJ...749...53R} {749, 53}

\bibitem[\protect\citeauthoryear{{Salim} et~al.,}{{Salim}
  et~al.}{2016}]{Salim2016}
{Salim} S.,  et~al., 2016, \mn@doi [\apjs] {10.3847/0067-0049/227/1/2}, \href
  {https://ui.adsabs.harvard.edu/abs/2016ApJS..227....2S} {227, 2}

\bibitem[\protect\citeauthoryear{{Shibuya}, {Ouchi}  \& {Harikane}}{{Shibuya}
  et~al.}{2015}]{Shibuya2015}
{Shibuya} T.,  {Ouchi} M.,   {Harikane} Y.,  2015, \mn@doi [\apjs]
  {10.1088/0067-0049/219/2/15}, \href
  {http://adsabs.harvard.edu/abs/2015ApJS..219...15S} {219, 15}

\bibitem[\protect\citeauthoryear{{Springel} \& {Hernquist}}{{Springel} \&
  {Hernquist}}{2005}]{Springel2005}
{Springel} V.,  {Hernquist} L.,  2005, \mn@doi [\apjl] {10.1086/429486}, \href
  {http://adsabs.harvard.edu/abs/2005ApJ...622L...9S} {622, L9}

\bibitem[\protect\citeauthoryear{{Toomre}}{{Toomre}}{1977}]{Toomre1977}
{Toomre} A.,  1977, in {Tinsley} B.~M.,  {Larson} D.~Campbell R.~B.~G.,  eds,
  Evolution of Galaxies and Stellar Populations. p.~401

\bibitem[\protect\citeauthoryear{{Welker}, {Dubois}, {Pichon}, {Devriendt}  \&
  {Chisari}}{{Welker} et~al.}{2018}]{Welker2018}
{Welker} C.,  {Dubois} Y.,  {Pichon} C.,  {Devriendt} J.,   {Chisari} N.~E.,
  2018, \mn@doi [\aap] {10.1051/0004-6361/201629007}, \href
  {http://cdsads.u-strasbg.fr/abs/2018A%26A...613A...4W} {613, A4}

\bibitem[\protect\citeauthoryear{{Wright} et~al.,}{{Wright}
  et~al.}{2017}]{Wright2017}
{Wright} A.~H.,  et~al., 2017, \mn@doi [\mnras] {10.1093/mnras/stx1149}, \href
  {https://ui.adsabs.harvard.edu/abs/2017MNRAS.470..283W} {470, 283}

\bibitem[\protect\citeauthoryear{{Wuyts} et~al.,}{{Wuyts}
  et~al.}{2011}]{Wuyts2011}
{Wuyts} S.,  et~al., 2011, \mn@doi [\apj] {10.1088/0004-637X/742/2/96}, \href
  {http://adsabs.harvard.edu/abs/2011ApJ...742...96W} {742, 96}

\bibitem[\protect\citeauthoryear{{Yang}, {Mo}, {van den Bosch}, {Pasquali},
  {Li}  \& {Barden}}{{Yang} et~al.}{2007}]{Yang2007}
{Yang} X.,  {Mo} H.~J.,  {van den Bosch} F.~C.,  {Pasquali} A.,  {Li} C.,
  {Barden} M.,  2007, \mn@doi [\apj] {10.1086/522027}, \href
  {https://ui.adsabs.harvard.edu/abs/2007ApJ...671..153Y} {671, 153}

\bibitem[\protect\citeauthoryear{{York} et~al.,}{{York}
  et~al.}{2000}]{York2000}
{York} D.~G.,  et~al., 2000, \mn@doi [\aj] {10.1086/301513}, \href
  {https://ui.adsabs.harvard.edu/abs/2000AJ....120.1579Y} {120, 1579}

\bibitem[\protect\citeauthoryear{{Zeng}, {Wang}  \& {Gao}}{{Zeng}
  et~al.}{2021}]{Zeng2021}
{Zeng} G.,  {Wang} L.,   {Gao} L.,  2021, \mn@doi [\mnras]
  {10.1093/mnras/stab2294}, \href
  {https://ui.adsabs.harvard.edu/abs/2021MNRAS.507.3301Z} {507, 3301}

\bibitem[\protect\citeauthoryear{{Zhang}, {Li}, {Kauffmann}, {Zou},
  {Catinella}, {Shen}, {Guo}  \& {Chang}}{{Zhang} et~al.}{2009}]{Zhang2009}
{Zhang} W.,  {Li} C.,  {Kauffmann} G.,  {Zou} H.,  {Catinella} B.,  {Shen} S.,
  {Guo} Q.,   {Chang} R.,  2009, \mn@doi [\mnras]
  {10.1111/j.1365-2966.2009.15050.x}, \href
  {https://ui.adsabs.harvard.edu/abs/2009MNRAS.397.1243Z} {397, 1243}

\makeatother
\end{thebibliography}

\bsp	% typesetting comment
\label{lastpage}
\end{document}